\documentclass[gmd, manuscript]{copernicus}

\usepackage{bm}       
\usepackage{listings} 
\usepackage{subcaption}

\lstdefinelanguage{Julia}{
  keywords={function, end, do, for, if, else, return, let, in, begin},
  sensitive=true,
  comment=[l]{\#},
  morestring=[b]",
}
\nolinenumbers
\begin{document}

\title{LCS.jl: A High-Performance, Multi-Platform Computational Model in Julia for
  Turbulent Particle-Laden Flows}

\Author[1]{Taketo}{Tominaga}
\Author[2]{Ryo}{Onishi}

\affil[1]{Engineering School, Department of Mechanical Engineering, Institute of Science Tokyo}
\affil[2]{Supercomputing Research Center, Institute of Integrated Research, Institute of Science Tokyo}

\correspondence{Taketo Tominaga (tominaga.t.2d38@m.isct.ac.jp)}

\runningtitle{LCS.jl: Multi-Platform DNS for Turbulent Particle-Laden Flows}

\runningauthor{Tominaga and Onishi}

\received{}
\pubdiscuss{}
\revised{}
\accepted{}
\published{}

\firstpage{1}

\maketitle

\begin{abstract}
Multiphase turbulent flow phenomena are observed not only in industrial devices but also
in environmental flows, and direct numerical simulation (DNS) plays a key role in their
investigation.
Many numerical models have been developed; nevertheless, few models are highly optimized
for GPU platforms, which represent the current mainstream in high-performance computing
(HPC).
In this study, we developed LCS.jl (Lagrangian Cloud Simulator in Julia), a
single-source and multi-platform multiphase turbulence simulation model implemented in
Julia language and KernelAbstractions.jl.
Validation results confirmed that the present fluid and particle statistics agree well
with those obtained in prior studies.
A GPU-native particle communication algorithm based on prefix-scan reduced the particle
communication cost from approximately 78\% (CPU-delegated) to 10\% of total execution
time.
LCS.jl achieved computational performance equivalent to the Fortran implementation in
many-processes computations.
For GPUs, strong scaling efficiency was maintained above 85\% (up to 256 GPUs) and weak
scaling efficiency above 90\% (up to 216 GPUs) on TSUBAME4.0 (a GPU supercomputer at
the Institute of Science Tokyo).
LCS.jl achieved a maximum speedup of 18.0$\times$ on GPUs over CPUs.
A trial heterogeneous execution achieved a 72\% reduction in execution time compared to
the CPU-only configuration even in configurations where the GPU was not the primary
compute device.
These results demonstrate that LCS.jl is a multiphase turbulence simulation platform
that achieves both portability and scalability across a variety of computational resource
configurations.
\end{abstract}


\introduction

Multiphase turbulent flow phenomena are observed not only in industrial devices but also
in environments.
In environments flows, one example is the growth process of cloud droplets in turbulent
clouds such as cumulus and cumulonimbus.

Understanding this process is relevant for reliable weather and climate modeling.
In turbulent flows, small inertial particles such as cloud droplets form clustering,
i.e., distribute non-uniformly in space.
It has been recognized that this clustering significantly affects the
collision-coalescence growth of cloud droplets.
Extensive DNS studies have been conducted to model the effect of
clustering~\citep{sundaram1997collision, wang2000statistical, ireland2016effect}.

Environmental flows are characterized by high Reynolds numbers.
Direct numerical simulations (DNSs) of such flows require spatial resolutions down to
the Kolmogorov scale~\citep{kolmogorov1941local, pope2000turbulent}.
The number of computational grid points increases rapidly with increasing Reynolds
number, requiring more computational cost.
In case of flow-particle multiphase simulations, for, e.g., droplet growth in turbulent
environment, the particle solver adds extra cost.
It is a kind of great challenge to conduct multiphase DNS for environmental studies even
on the latest HPC systems.
\citet{onishi2015lagrangian} and \citet{onishi2016reynolds} addressed this challenge and
clarified the Reynolds number dependence of collision kernel through large-scale
computations.
However, evaluation under even higher Reynolds number conditions remains computationally
prohibitive.
Many numerical models have been developed to enable high-Reynolds-number computations.
Nevertheless, few models are highly optimized for GPU platforms, which represent the
current mainstream in high-performance computing (HPC).
With anticipation of future architectural diversification, ensuring multi-platform
compatibility i.e., performance portability across different architectures is becoming
increasingly important.

The Euler--Lagrangian framework is a widely used approach for multiphase DNS that tracks
inertial particles in turbulent flows.
One implementation of this framework is the Lagrangian Cloud Simulator (LCS), a Fortran
model developed by \citet{onishi2015lagrangian}.
LCS adopts distributed-memory parallelization via MPI and achieved large-scale parallel
computation on HPC environments.
However, conventional multiphase DNS models in Fortran or C have been highly optimized
for multi-CPU environments.
Supporting multi-GPU architectures while ensuring high performance in both fluid and
particle solvers is a critical task for such models.

There are two challenges in GPU-porting such models.
The first one concerns the particle communication algorithm.
In fluid computation, the communication range can be determined statically by the domain
decomposition.
In particle computation, by contrast, the number and destination of particles migrating
across subdomain boundaries are changing in time, requiring dynamic communication.
Conventional CPU-based implementations handle this communication sequentially, which is
incompatible with GPU execution.
The second challenge concerns performance portability.
The development of a multi-platform model that achieves high performance on both CPUs
and GPUs is a non-trivial task.
In recent years, Julia has attracted attention in the HPC community as a language that
achieves C/Fortran-level execution performance and high expressiveness through
just-in-time (JIT) compilation and type
specialization~\citep{bezanson2017julia}.
With the development of GPU extensions~\citep{besard2019effective} and
vendor-agnostic kernel abstractions -- KernelAbstractions.jl
\citep{churavy2020kernelabstractions} --, single-source and multi-platform design has
become practical.
In other words, Julia has the potential to achieve high performance, high expressiveness,
and performance portability within a single language.
For fluid solvers alone, Julia implementations such as
Oceananigans.jl~\citep{ramadhan2020oceananigans} and
MPAS-Ocean~\citep{bishnu2023julia} have been reported to achieve performance comparable
to or exceeding that of Fortran.
However, only a limited number of Julia-based implementations of Euler--Lagrangian
multiphase DNS solvers have been reported~\citep{ramadhan2020oceananigans}.
However, to the authors’ knowledge, only a single Julia-based implementation of Euler--Lagrangian
multiphase DNS solvers has been reported~\citep{ramadhan2020oceananigans} and it is limited to tracer particles. No Julia-based multiphase DNS solver for inertial particles has been reported, for which particle communication costs would be more significant.

In this study, we aim to develop LCS.jl (Lagrangian Cloud Simulator in Julia), a
single-source and multi-platform multiphase DNS model implemented in Julia language and
KernelAbstractions.jl.
We validate the developed model through comparison of fluid and particle statistics with
prior studies.
We propose a GPU-native particle communication algorithm and evaluate its performance
against a CPU-delegated implementation.
We further evaluate the computational performance of the single-source design, including
strong- and weak-scaling in large-scale parallel computations, GPU-CPU performance
comparison using the identical codebase, and a heterogeneous execution configuration in
which GPUs are utilized in an auxiliary capacity.
Through these contributions, we assess the portability and scalability of LCS.jl across
a variety of computational resource configurations.


\section{Methods}

\subsection{Flow Phase}

The carrier fluid obeys the following governing equations:
\begin{align}
  \nabla \cdot U &= 0, \label{eq:continuity} \\
  \frac{\partial U}{\partial t} + (U \cdot \nabla) U
    &= -\frac{1}{\rho} \nabla p + \nu \nabla^2 U + F(\mathbf{x}, t), \label{eq:momentum}
\end{align}
where $U$ is the fluid velocity, $\rho$ is the density, $p$ is the pressure, and $F$ is
the external forcing.
The kinematic viscosity $\nu$ was set to
$1.5 \times 10^{-5}\,\mathrm{m}^2\,\mathrm{s}^{-1}$, corresponding to atmospheric
conditions (1013\,hPa, 298\,K).
Turbulence is sustained by the forcing $F$, which injects energy into large scales at
wavenumber $|k| < 2.5$.
For the turbulent forcing, the Reduced Communication Forcing
(RCF)~\citep{onishi2011largescale}, an improved large-scale forcing scheme, was adopted.
RCF reduces communication and computational costs by applying a box-mean filter to the
velocity field before transforming it to wavenumber space, rather than transforming the
entire field, while still confining energy injection to large scales.
The governing equations were discretized on a uniform Cartesian grid, with variables
stored in a marker and cell (MAC) arrangement~\citep{HarlowWelch1965}.
For spatial discretization, the conservative fourth-order central difference
scheme~\citep{morinishi1998fully} was applied to the convection term, fourth-order
central differences were applied to the viscous term, and second-order central
differences were applied to the pressure term.
Time integration was performed using a two-stage, second-order Runge--Kutta method
(RK2).
Pressure--velocity coupling was handled by the highly simplified marker and cell (HSMAC)
scheme~\citep{hirt1972calculating} with Red--Black coloring.
Iteration was continued until the root-mean-square (RMS) of the velocity divergence fell
below $\delta/\Delta$, where $\Delta$ is the grid spacing and $\delta$ was set to
$10^{-3}$ following \citet{onishi2015lagrangian}.
The time step $\Delta t$ was determined to satisfy both the convective and viscous CFL
constraints for stability:
\begin{align}
  \mathrm{CFL}_\mathrm{conv} &= \max \frac{\|U\| \Delta t}{\Delta} \leq C_1, \\
  \mathrm{CFL}_\mathrm{vis}  &= \frac{\nu \Delta t}{\Delta^2} \leq C_2,
\end{align}
where $C_1 = C_2 = 0.3$.
\citet{onishi2013efficient} demonstrated that the present finite-difference model
employing these numerical methods maintains sufficient accuracy compared to spectral
methods.
The computational domain is a cube of side $2\pi L_0$ with periodic boundary conditions
applied in all directions.
The grid spacing is $\Delta = 2\pi L_0 / N$, and the bulk Reynolds number is defined as
$Re = U_0 L_0 / \nu$, where $U_0$ is the reference velocity and $L_0$ is the reference
length.
Three-dimensional domain decomposition was used for inter-process parallelization.
The decomposition was chosen so that each subdomain is as close to cubic as possible,
minimizing the surface-area-to-volume ratio and reducing communication cost.

\subsection{Particle Phase}

Particle motion obeys the following system of ordinary differential equations:
\begin{align}
  \frac{\mathrm{d}\bm{X}}{\mathrm{d}t} &= \bm{V}, \\
  \frac{\mathrm{d}\bm{V}}{\mathrm{d}t} &= -\frac{f}{\tau_p} \bigl(\bm{V} - U(\bm{X}, t)\bigr),
\end{align}
where $\bm{X}$ and $\bm{V}$ are the particle position and velocity, and $U(\bm{X},t)$
is the fluid velocity at the particle position.
$\tau_p = \frac{2}{9} (\rho_p/\rho) (r^2/\nu)$ is the relaxation time of a particle
with radius $r$ and density $\rho_p$.
The particle-to-fluid density ratio $\rho_p/\rho$ was set to $8.43 \times 10^2$,
corresponding to water under atmospheric conditions (1013\,hPa, 298\,K).
The non-linear drag correction factor $f$ and the particle Reynolds number $Re_p$ are
defined as follows~\citep{rowehenwood1961}:
\begin{align}
  f    &= 1 + 0.15\,Re_p^{0.687}, \\
  Re_p &= \frac{2r |\bm{V} - U(\bm{X})|}{\nu}.
\end{align}
The fluid velocity $U(\bm{X})$ at the particle position $\bm{X}$ was computed by
trilinear interpolation from the eight surrounding cell-center values.
Time integration was performed using a two-stage, second-order Runge--Kutta method.

\subsection{Implementation and Optimization in Julia}

\subsubsection{Performance Portability: Single Source, Multi-Platform Strategy}

LCS.jl is designed to achieve performance portability across arbitrary platforms
including CPUs and GPUs with a single source code.
The core of this design is the KernelAbstractions.jl~\citep{churavy2020kernelabstractions}
library.
In the present model, abstraction functions at an even higher level than this library
were adopted (Fig.~\ref{fig:for-loop-comparison}).
\texttt{Parallel.foraxes} is a for-loop abstraction provided by Parallel, a custom
parallel execution module in LCS.jl, and generates platform-specific code for a variety
of computational platforms from a single source.
An explicit data movement design was also adopted.
Compared to directive-based GPU porting approaches such as OpenACC, this suppresses the
proliferation of pragma annotations required for performance optimization and maintains
code readability.

\begin{figure}[t]
\begin{lstlisting}[language=Julia]
# N: Problem size
# U: Fluid velocity field
# dUdt: Time derivative of U
# dx: Grid spacing
# backend: Target backend (e.g., CPU, CUDA, AMD, Metal, OpenAPI, etc.)

Parallel.foraxes(
  backend, (2:N-1, 2:N-1, 2:N-1))
) do i, j, k
  dUdt[i, j, k] = - (U[i+1, j, k] - U[i-1, j, k]) / (2 * dx)
end
\end{lstlisting}
(a)
\begin{lstlisting}[language=Fortran]
do k = 2, N-1
  do j = 2, N-1
    do i = 2, N-1
      dUdt(i, j, k) = - (U(i+1, j, k) - U(i-1, j, k)) / (2 * dx)
    end do
  end do
end do
\end{lstlisting}
(b)
\caption{Implementation of a finite-difference kernel in LCS.jl (a) and Fortran (b).}
\label{fig:for-loop-comparison}
\end{figure}

\subsubsection{HALO Communication}

In LCS.jl, communication--computation overlap and time-blocking were introduced as
communication optimizations in the HSMAC iteration loop, which occupies the largest
fraction of execution time.
Communication--computation overlap hides communication by concurrently performing HALO
region communication and computation that does not depend on the HALO region.
HALO region communication refers to exchanging boundary data between neighboring
processes in distributed computing.
Time-blocking reduces the number of communication calls by aggregating multiple iteration
steps and exchanging HALO regions in a single batch.
The pressure correction in the HSMAC method uses second-order central differences,
requiring a HALO width of one cell.
This is smaller than the HALO width required for the fourth-order convection
discretization (three cells).
By grouping three HSMAC iteration steps together and exchanging a three-cell HALO in a
single communication call, the number of communications is reduced by a factor of three
without increasing memory usage.

\subsubsection{Parallel Particle Communication Based on Prefix-Scan}

In particle tracking, the communication destination of particles that cross subdomain
boundaries is determined only at runtime.
Unlike the static communication in HALO exchange, dynamic communication is therefore
required.

In the original LCS (Fortran implementation), particle communication was implemented
sequentially.
All particles are scanned in order; each time a particle outside the subdomain boundary
is detected, a per-direction counter is incremented to determine its storage position in
the send buffer.
In this scheme, the storage position of particle $i$ depends on the processing result of
particles $1$ through $i-1$ (i.e., the counter value).
This sequential dependency makes the scheme incompatible with GPU environments where
multiple threads update the same counter concurrently.

In LCS.jl, a GPU-native particle communication algorithm based on prefix-scan was
implemented.
The procedure consists of the following three stages.
\begin{enumerate}
  \item \textbf{Mask computation}: A boundary-crossing check is performed for all
    particles to generate a \texttt{mask} array. Particles to be sent are assigned~1;
    particles remaining within the subdomain are assigned~0. Each particle's check is
    independent, so this stage is executed in parallel by GPU threads.
  \item \textbf{Prefix-scan}: The cumulative sum (prefix-scan) of the mask array is
    computed to determine the storage position of each particle in the send buffer in one
    pass. The storage position of particle $i$ is given by
    $\mathrm{scan}[i] = \sum_{j=1}^{i} \mathrm{mask}[j]$.
  \item \textbf{Parallel packing}: For particles with mask~$= 1$, each particle is
    written in parallel by GPU threads to the position indicated by $\mathrm{scan}[i]$.
    Particles with mask~$= 0$ are excluded from writing, so no threads overwrite to the
    same destination and no data race occurs.
\end{enumerate}
As a concrete example, consider five particles where particles 2 and 4 are to be sent:
\begin{align*}
  \mathrm{mask}: & \quad 0,\ 1,\ 0,\ 1,\ 0 \\
  \mathrm{scan}: & \quad 0,\ 1,\ 1,\ 2,\ 2
\end{align*}
Particle~2 is written to $\mathrm{buffer}[1]$ via $\mathrm{scan}[2] = 1$, and
particle~4 is written to $\mathrm{buffer}[2]$ via $\mathrm{scan}[4] = 2$.

In the sequential implementation, storage positions are determined one particle at a
time by incrementing a counter, while the next particle cannot be processed until the
previous one is complete.
In the present prefix-scan implementation, both position determination and packing were
executed in parallel by GPU threads.
This eliminates the sequential dependency and enables efficient GPU-native particle
communication.


\section{Results}

\subsection{Validation}

\subsubsection{Flow Phase}

\begin{table}[b]
\caption{Flow statistics in the statistically-steady state at different resolutions.
  $-S$ denotes the skewness of the velocity gradient; $F$ denotes the flatness.}
\label{tab:steady-state-stats}
\begin{tabular}{cccccc}
\tophline
$N^3$      & $u_\mathrm{rms}$ & $Re_\lambda$ & $k_\mathrm{max}l_\eta$ & $-S$ & $F$ \\
\middlehline
$128^3$    & 0.966 & 79.3 & 2.09 & 0.482 & 4.82 \\
$256^3$    & 0.993 & 127  & 2.05 & 0.488 & 5.45 \\
$512^3$    & 1.03  & 209  & 2.02 & 0.564 & 6.58 \\
$1024^3$   & 1.00  & 333  & 2.14 & 0.570 & 7.47 \\
$2048^3$   & 0.999 & 536  & 2.16 & 0.604 & 9.02 \\
\bottomhline
\end{tabular}
\belowtable{}
\end{table}

\begin{figure}[b]
\includegraphics[width=0.75\textwidth]{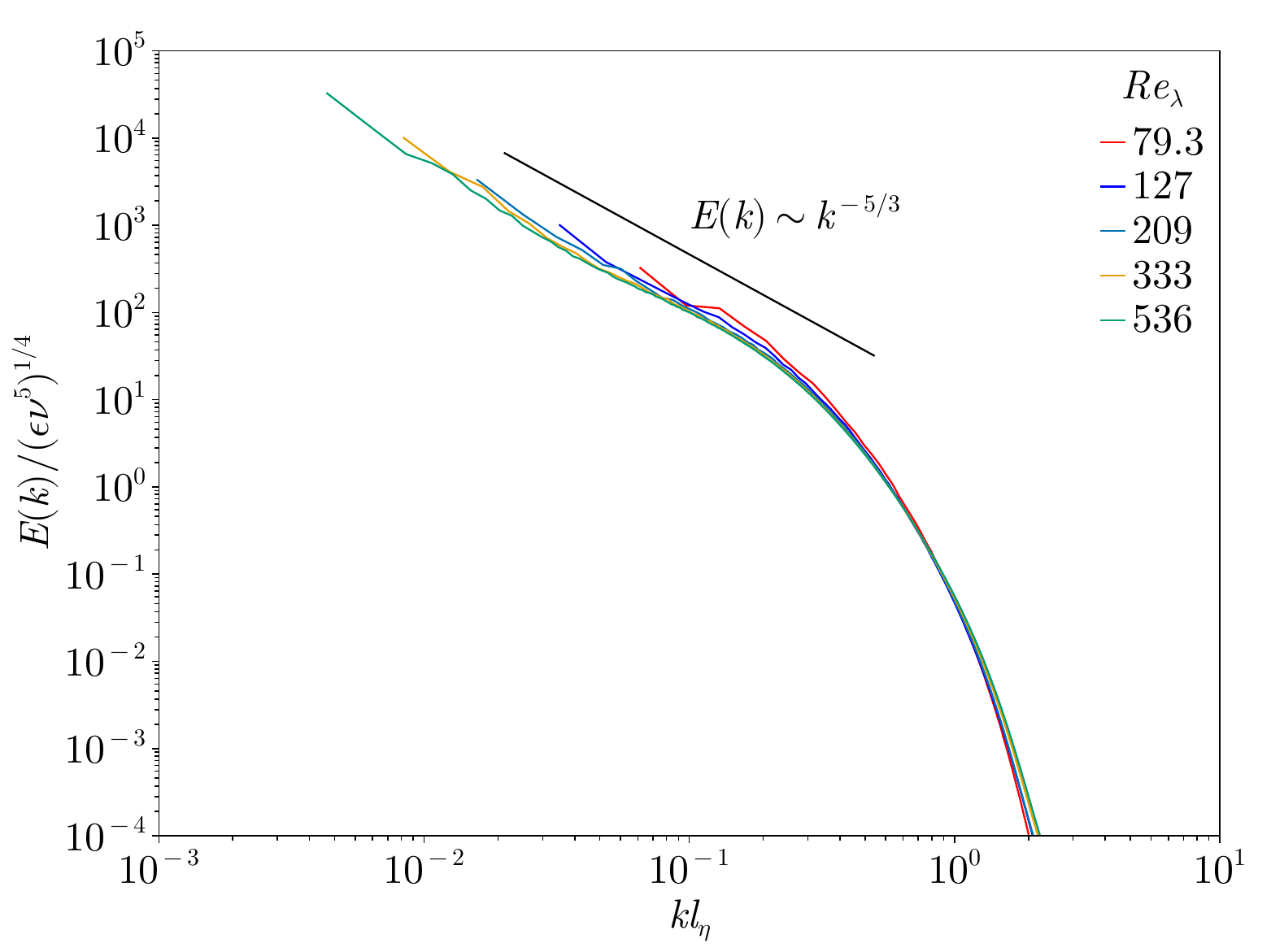}
\caption{Energy spectra $E(k)$ normalized by the Kolmogorov velocity scale
  $(\varepsilon\nu^5)^{1/4}$, where $\varepsilon$ is the mean energy dissipation rate
  and $\nu$ is the kinematic viscosity, as a function of the normalized wavenumber
  $kl_\eta$, where $l_\eta = (\nu^3/\varepsilon)^{1/4}$ is the Kolmogorov length scale.
  Results are shown for resolutions $N^3 = 128^3$ through $2048^3$
  ($Re_\lambda \approx 79.3$ to $536$).
  The dashed line indicates the Kolmogorov $k^{-5/3}$ scaling.}
\label{fig:kolmogorov-scaling}
\end{figure}

Table~\ref{tab:steady-state-stats} shows the statistically-steady state statistics
obtained from GPU computations using LCS.jl.
Computations were performed up to $N^3 = 2048^3$.
All cases satisfy $k_\mathrm{max}\eta \approx 2$, where $k_\mathrm{max} = N/2$ is the
maximum wavenumber and $\eta$ is the Kolmogorov length scale, ensuring sufficient
spatial resolution down to the Kolmogorov scale.
The obtained statistics, including root-mean-square velocity $u_\mathrm{rms}$,
Taylor-scale based Reynolds number $Re_\lambda$, skewness of velocity derivative $-S$,
and flatness of velocity derivative $F$, agreed well with the prior
study~\citep{onishi2011largescale}.
Figure~\ref{fig:kolmogorov-scaling} shows kinetic energy spectra $E(k)$, defined as
$E(k) = \frac{1}{2} \sum'_k |\hat{u}(\bm{k})|^2$, where $\hat{\cdot}$ denotes the
Fourier transform,
$\sum'_k \equiv \frac{4\pi k^2}{N_k} \sum_{k-1/2 \leq |\bm{k}| < k+1/2}$,
and $N_k$ is the number of wavevectors $\bm{k}$ satisfying
$k - 1/2 \leq |\bm{k}| < k + 1/2$.
For the largest case with $Re_\lambda \approx 536$, an inertial subrange spanning two
decades is confirmed, demonstrating that the energy cascade is properly captured.
This confirmed that LCS.jl correctly reproduced turbulence statistics, validating the
fluid solver implementation.

\subsubsection{Particle Phase}\label{sec:particle-phase-validation}

\begin{figure}[b]
\includegraphics[width=0.60\textwidth]{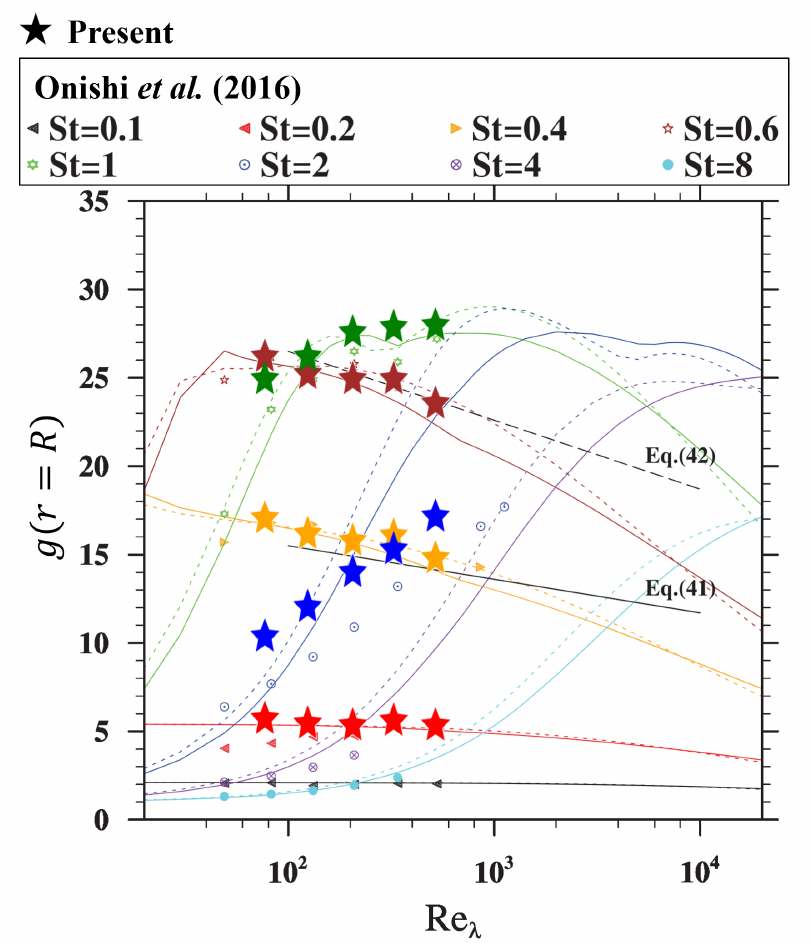}
\caption{Radial distribution function at contact $g(r=R)$ as a function of the
  Taylor-microscale Reynolds number $Re_\lambda$.
  Stars denote the present results ($\mathrm{St} = 0.2$ to $2.0$); other symbols denote
  the results ($\mathrm{St} = 0.1$ to $8$) of Onishi et al.\ (2016).}
\label{fig:gr-validation}
\end{figure}

Figure~\ref{fig:gr-validation} shows the dependence of the radial distribution function
at contact $g(r=R)$ on Stokes number $\mathrm{St}$ and $Re_\lambda$, where $\mathrm{St}$
is the ratio of particle relaxation time $\tau_p$ to Kolmogorov timescale.
The quantity $g(r=R)$ quantifies the clustering intensity, where unity indicates a
uniform distribution.
This quantity is used in the calculation of collision kernels in cloud microphysics.
The results agreed well with the prior study~\citep{onishi2016reynolds} across all
$\mathrm{St}$ and $Re_\lambda$ conditions.
This confirmed that LCS.jl quantitatively reproduced the $\mathrm{St}$ and $Re_\lambda$
dependence of particle clustering in turbulence, validating the particle tracking
implementation.

\subsection{Computational Performance}

\begin{table}[b]
\caption{TSUBAME4.0 node configuration.}
\label{tab:tsubame4-hardware}
\begin{tabular}{ll}
\tophline
\textbf{Component}       & \textbf{Specification} \\
\middlehline
CPU                      & AMD EPYC 9654 2.4\,GHz $\times$ 2 Socket \\
Cores / Threads          & 96 cores / 192 threads $\times$ 2 Socket \\
Memory                   & 768\,GiB (DDR5-4800) \\
GPU                      & NVIDIA H100 SXM5 94\,GB HBM2e $\times$ 4 \\
CPU--GPU Interconnect    & PCIe Gen5 $\times$16 (64\,GB/s one-way) \\
SSD                      & 1.92\,TB NVMe U.2 SSD \\
Network Interconnect     & InfiniBand NDR200 200\,Gbps $\times$ 4 \\
\bottomhline
\end{tabular}
\belowtable{}
\end{table}

Computational performance was evaluated on TSUBAME4.0 at the Institute of Science Tokyo.
Table~\ref{tab:tsubame4-hardware} shows the node configuration.

\subsubsection{HALO Communication Optimization}\label{sec:halo-comm-opt}

\begin{figure}[b]
\includegraphics[width=0.75\textwidth]{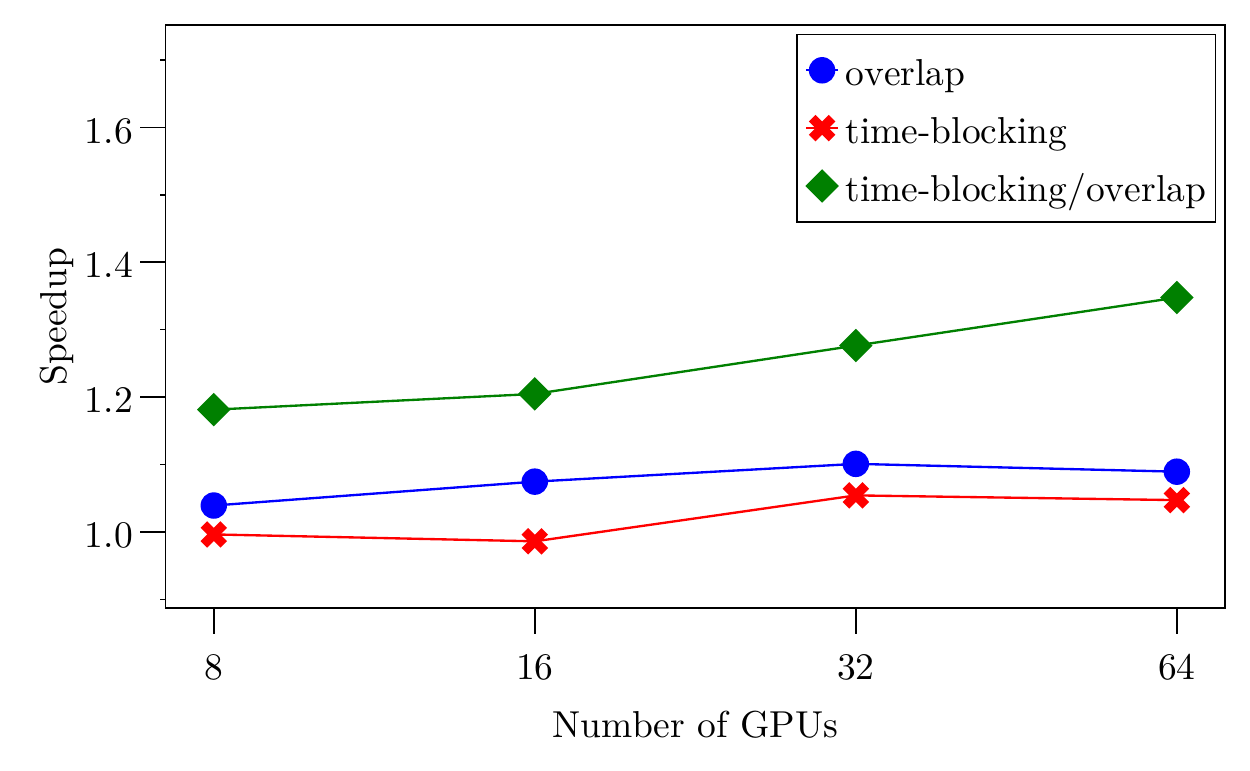}
\caption{Speedup of wall-clock time per time-integration step as a function of the number of GPUs
  ($N^3 = 1500^3$, $N_p = 750^3$), for three HALO communication optimizations:
  communication--computation overlap, time-blocking, and their combination.
  The speedup is defined as the ratio of execution time relative to the baseline with no
  optimization applied.}
\label{fig:halo-comm-opt}
\end{figure}

Figure~\ref{fig:halo-comm-opt} shows the speedup for HALO communication optimizations
with respect to the number of GPUs.
Measurements were performed for grid size $1500^3$ and particle count $750^3$, ranging
from 8 to 64 GPUs.
The speedup was defined as the ratio of execution time relative to the baseline with no
optimization applied.
The combined use of the time-blocking method and communication--computation overlap
yielded the highest speedup of $1.34\times$ at 64 GPUs.
These results confirmed the performance improvement obtained by communication
optimization.

\subsubsection{Particle Communication Optimization Based on Prefix-Scan}\label{sec:particle-comm-opt}

Table~\ref{tab:particle-comm} shows the particle communication cost for two
implementations: a GPU-native implementation based on prefix scan and a CPU-delegated
sequential implementation.
The measurement conditions were grid size $N^3 = 1500^3$, particle count $N_p = 750^3$,
and 8 GPUs.
The particle arrays consisted of velocity fields and positions for the RK2 stages (2
each), velocity time derivatives (1), particle IDs, and particle diameters (1 each).
Since vector quantities had 3 components, the total number of Float64 arrays was
$(2+2+1) \times 3 + 2 = 17$.
The number of particles per rank was
$N_p^\mathrm{rank} = 750^3 / 8 \approx 5.3 \times 10^7$,
and the data transfer volume was
\begin{equation}
  17 \times 8\,\mathrm{B} \times 5.3 \times 10^7 \approx 7.2\,\mathrm{GB}.
\end{equation}
Using the PCIe Gen5 $\times$16 one-way bandwidth of 64\,GB/s on TSUBAME4.0, the
estimated transfer time for each of D2H (device-to-host) and H2D (host-to-device) is
\begin{equation}
  t_\mathrm{D2H} = t_\mathrm{H2D} = 7.2\,\mathrm{GB} / 64\,\mathrm{GB/s}
    \approx 113\,\mathrm{ms}.
\end{equation}
Sequential packing on the CPU was limited by the single-thread memory bandwidth.
The single-core memory bandwidth was measured using the STREAM Triad
benchmark~\citep{mccalpin1991stream} and confirmed to be approximately 41\,GB/s.
The estimated cost for each of packing and unpacking was
\begin{equation}
  t_\mathrm{pack} = t_\mathrm{unpack} = 7.2\,\mathrm{GB} / 41\,\mathrm{GB/s}
    \approx 176\,\mathrm{ms}.
\end{equation}
In the GPU-native implementation, D2H/H2D transfer of the full particle arrays is not
required.
The measured packing and unpacking times were 12\,ms and 60\,ms, respectively.
The communication cost per RK2 stage in the GPU-native implementation based on
prefix-scan was $12\,\mathrm{ms} + 60\,\mathrm{ms} = 72\,\mathrm{ms}$, which was
approximately
$578\,\mathrm{ms} / 72\,\mathrm{ms} \approx 8.0$ times faster than the theoretical
lower bound of the CPU-delegated approach,
$(113\,\mathrm{ms} + 176\,\mathrm{ms}) \times 2 = 578\,\mathrm{ms}$.
The higher unpacking cost (60\,ms vs.\ 12\,ms) is attributed to the compaction step, in
which invalid particles are moved to the end of the array via intermediate buffer copies
in GPU memory.
Note that this estimate excluded the computational cost of packing itself; the actual
cost of CPU delegation is therefore even higher.

The impact of this communication speedup on overall performance was evaluated.
The wall-clock time per time-integration step under the measurement conditions of 8 GPUs,
$N^3 = 1500^3$, and $N_p = 750^3$ was 1490\,ms
(Sect.~\ref{sec:scaling}).
With CPU delegation, the particle communication cost for the two RK2 stages accounted
for approximately 78\% of the total execution time
($578\,\mathrm{ms} \times 2 / 1490\,\mathrm{ms} \approx 0.78$).
The GPU-native implementation reduced this fraction to approximately 10\%
($0.78 / 8.0 \approx 0.098$).

This reduction eliminates the communication bottleneck that would otherwise negate the
computational advantage of GPU execution, enabling effective utilization of GPU
performance in large-scale multiphase flow simulations.

\begin{table}[t]
\caption{Particle communication cost per RK2 stage for the GPU-native (prefix-scan,
  measured) and CPU-delegated sequential (estimated) implementations
  ($N^3 = 1500^3$, $N_p = 750^3$, 8 GPUs).
  MPI communication cost is equivalent in both implementations
  ($<$1\,ms per direction) and is omitted.
  The CPU-delegated estimates use PCIe Gen5 $\times$16 one-way bandwidth
  ${\sim}64$\,GB/s and EPYC~9654 single-core bandwidth ${\sim}41$\,GB/s.}
\label{tab:particle-comm}
\begin{tabular}{lcc}
\tophline
\textbf{Phase}
  & \textbf{GPU-native (measured)}
  & \textbf{CPU-delegated sequential (estimated)} \\
\middlehline
D2H transfer (full particle arrays) & ---          & ${\sim}$113\,ms \\
Packing                             & 12\,ms       & ${\sim}$176\,ms \\
Unpacking                           & 60\,ms       & ${\sim}$176\,ms \\
H2D transfer (full particle arrays) & ---          & ${\sim}$113\,ms \\
\textbf{Total (per stage)}          & \textbf{72\,ms} & $\mathbf{{\sim}578\,\mathrm{ms}}$ \\
\bottomhline
\end{tabular}
\belowtable{}
\end{table}

\subsubsection{Julia vs Fortran Comparison}\label{sec:julia-vs-fortran}

\begin{figure}[t]
\includegraphics[width=0.75\textwidth]{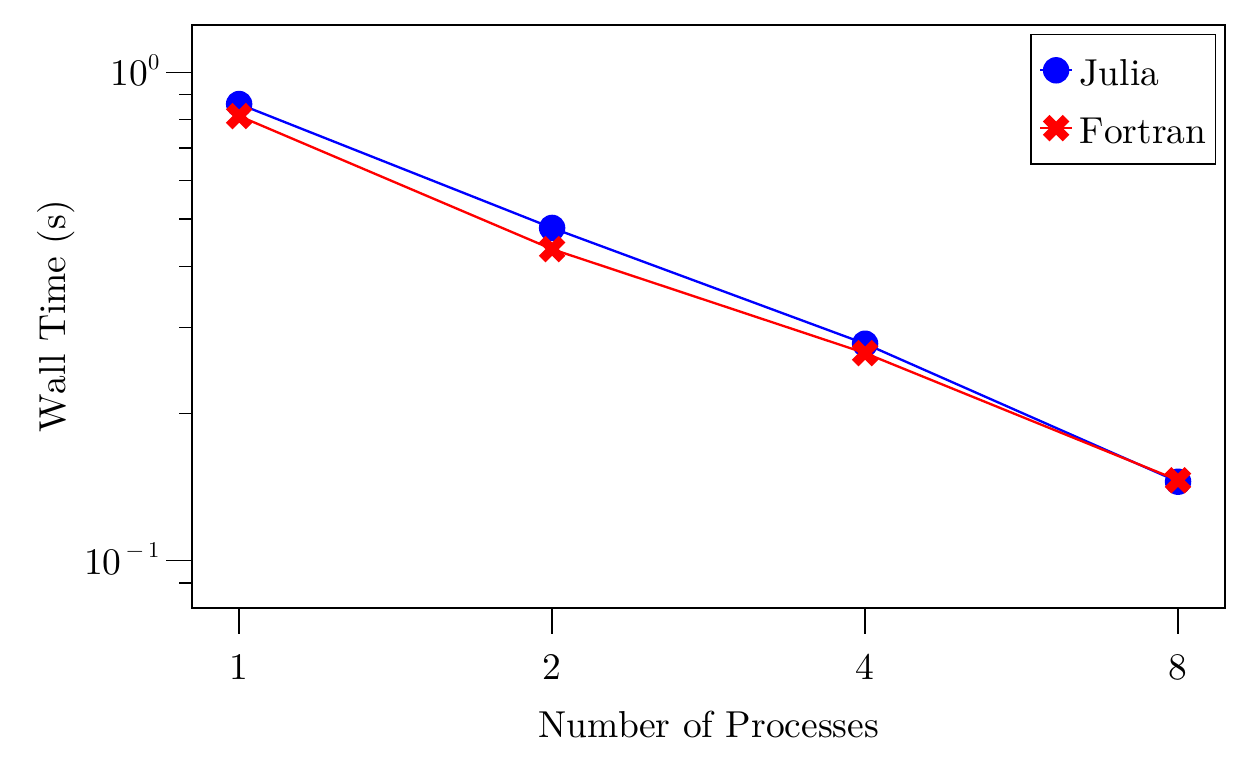}
\caption{Wall-clock time per time-integration step as a function of the number of CPU processes for
  LCS.jl (Julia) and the Fortran implementation ($N^3 = 256^3$, $N_p = 128^3$).}
\label{fig:julia-vs-fortran}
\end{figure}

Figure~\ref{fig:julia-vs-fortran} shows wall-clock time per time-integration step as a function of the
number of CPU processes for LCS.jl (Julia) and the Fortran implementation.
Computations were performed with grid size $N^3 = 256^3$ and particle count
$N_p = 128^3$, with identical computational settings in both implementations.
Execution time was measured as a function of the number of processes to evaluate
parallel performance.

At 1, 2, and 4 processes, LCS.jl showed an execution time up to approximately 10\%
longer than the Fortran implementation.
At 8 processes, both implementations exhibited comparable execution times.

At small process counts, the contribution of communication cost was small, and
differences in raw computational performance were dominant.
LCS.jl uses identical kernel code for both CPU and GPU.
The Fortran implementation applies CPU-specific loop transformations, such as variable
precomputation and loop fusion.
In a single kernel targeting both CPU and GPU execution, constraints on inter-thread
data sharing prevent equivalent optimizations from being applied in full.
This difference manifests as a performance gap of up to approximately 10\% at small
process counts.
However, as the number of processes increases, where the contribution of communication
cost grows, the performance became comparable to the Fortran implementation at 8
processes due to communication optimizations applied
(Sect.~\ref{sec:halo-comm-opt}).

These results confirmed that LCS.jl achieves computational performance equivalent to the
Fortran implementation in computations using many processes.
This demonstrated that the single-source and multi-platform design does not sacrifice
computational performance.

\subsubsection{Scaling}\label{sec:scaling}

Scaling measurements were performed on TSUBAME4.0 using up to 64 nodes.
The baseline problem size was set to $750^3$ grids and $325^3$ particles, which is the
maximum problem size that fits within a single GPU memory.
The particle count was set to range from 1/10 to 10 times the grid count, which is the
range used in typical cloud microphysics computations.
Performance evaluation was conducted independently of physical statistical analysis.
To isolate scaling performance from physical transients, the number of Poisson
iterations was fixed at 6, which is the typical value observed at statistically steady
state.

Julia employs just-in-time (JIT) compilation, which compiles code on its first
execution.
To exclude the effect of JIT compilation, the first 10 steps were treated as warmup and
excluded from the measurements.

To compare GPU and CPU under identical conditions, one device (1 CPU / 1 GPU) was
assigned per process.
However, for CPUs, assigning 1 process to 1 CPU (96 threads) is not optimal.
In memory-bound computations, a high thread count reduces memory bandwidth efficiency.
To determine the optimal balance, performance measurements were conducted with the total
thread count fixed at 384 (4 CPUs $\times$ 96 threads) with $N^3 = 1500^3$ grids and
$N_p = 750^3$ particles, while the number of threads per process was varied.
A maximum speedup of $2.7\times$ was achieved at 4 threads per process over the 1 CPU
(96 thread) per process baseline.
This optimization efficiency of 2.7 will be used to estimate the effective performance
on CPUs when comparing performances on CPUs and GPUs.

\begin{figure}[t]
\includegraphics[width=0.80\textwidth]{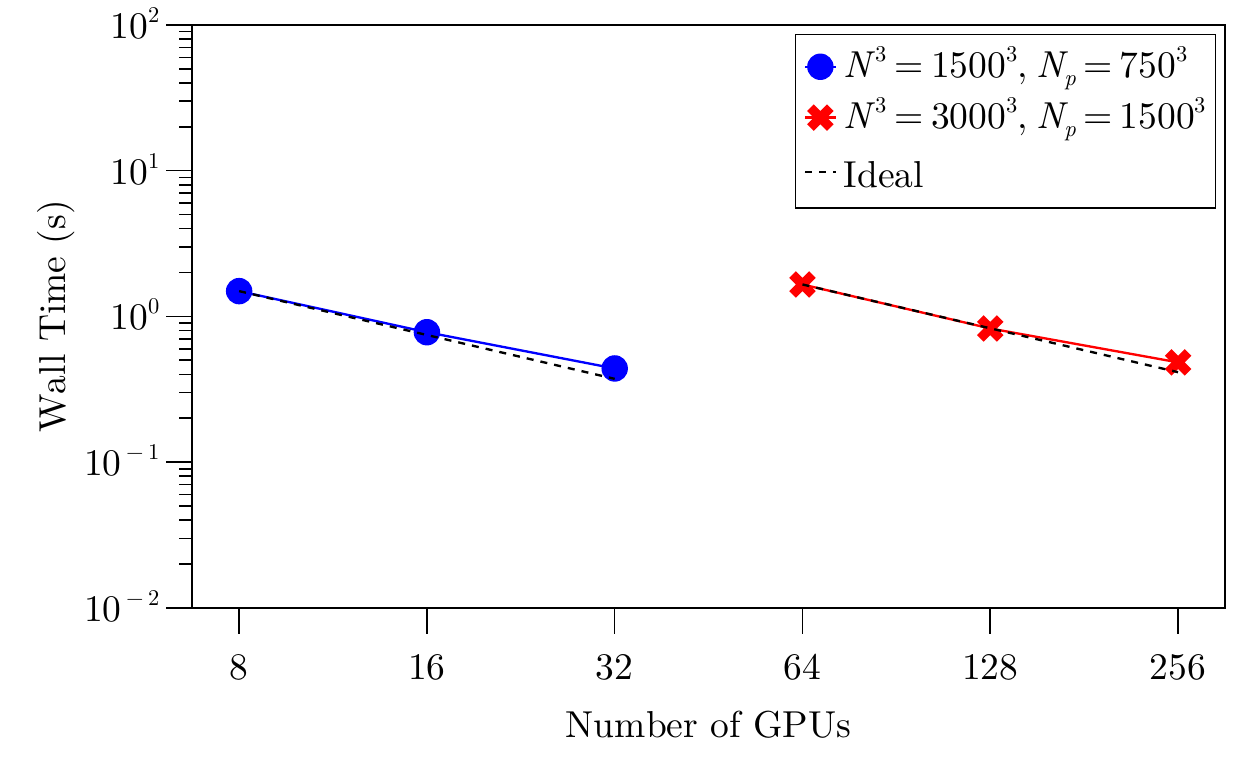}\\
(a)\\[0.5em]
\includegraphics[width=0.80\textwidth]{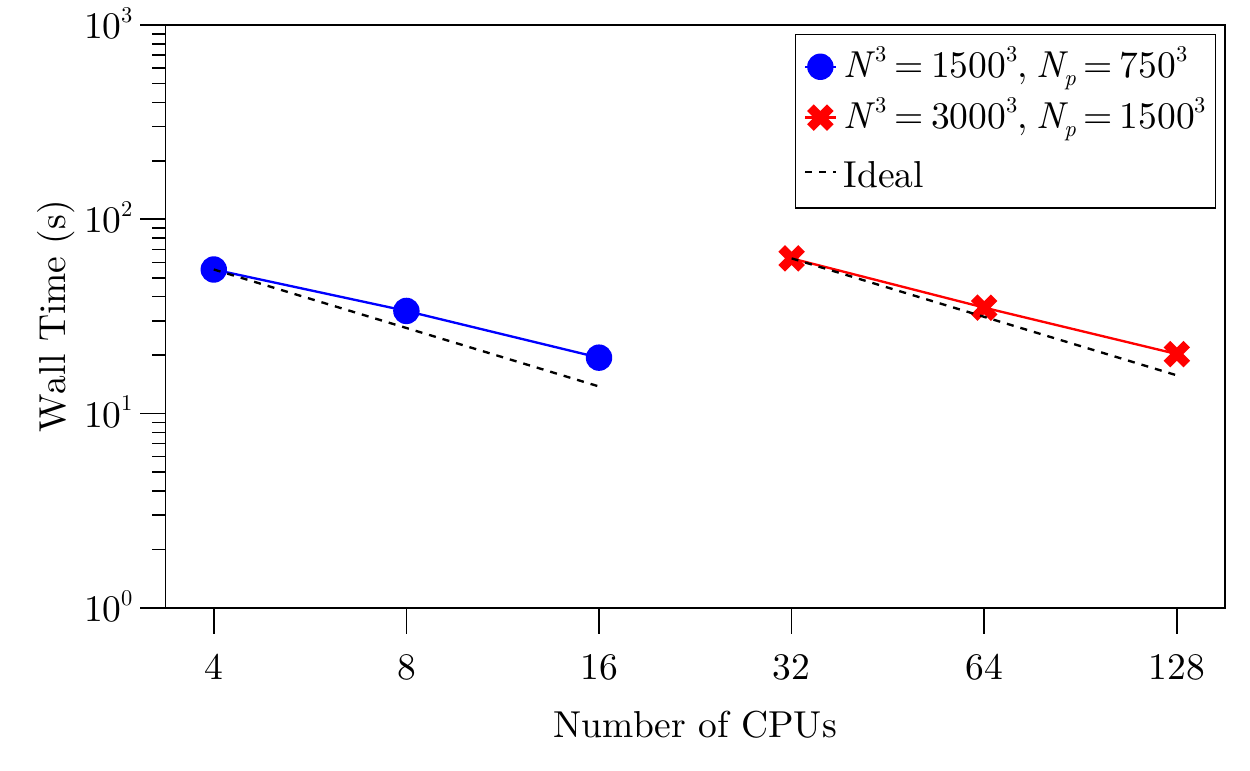}\\
(b)
\caption{Wall-clock time per time-integration step as a function of the number of devices for fixed total
  problem sizes. (a) GPU execution; (b) CPU execution.
  The black line indicates ideal strong scaling.}
\label{fig:strong-scaling}
\end{figure}

Figure~\ref{fig:strong-scaling} shows strong scaling results for GPU and CPU.
Total problem sizes of $N^3 = 1500^3$ and $3000^3$ grids were used.
Since TSUBAME4.0 has 4 GPUs and 2 CPUs per node, scaling was measured up to 256 GPUs
and 128 CPUs due to node count constraints.
Strong scaling efficiency is defined as $E_s = T_0 / (N T_N)$, where $T_0$ is the
execution time at the baseline device count and $T_N$ is the execution time at $N$
devices.
The baseline device count was 8 GPUs / 4 CPUs for $N^3 = 1500^3$, and 64 GPUs / 32
CPUs for $N^3 = 3000^3$.

Figure~\ref{fig:strong-scaling} confirms that strong scaling efficiency was maintained
above 85\% for GPUs (up to 256) and above 70\% for CPUs (up to 128).
The lower strong scaling efficiency of CPUs (70\%) compared to GPUs can be attributed to
memory bandwidth contention among threads within a single CPU socket, not to a
limitation of the multi-platform design.
This contention can be mitigated by adjusting the process and process-thread balance.

\begin{figure}[b]
\includegraphics[width=0.80\textwidth]{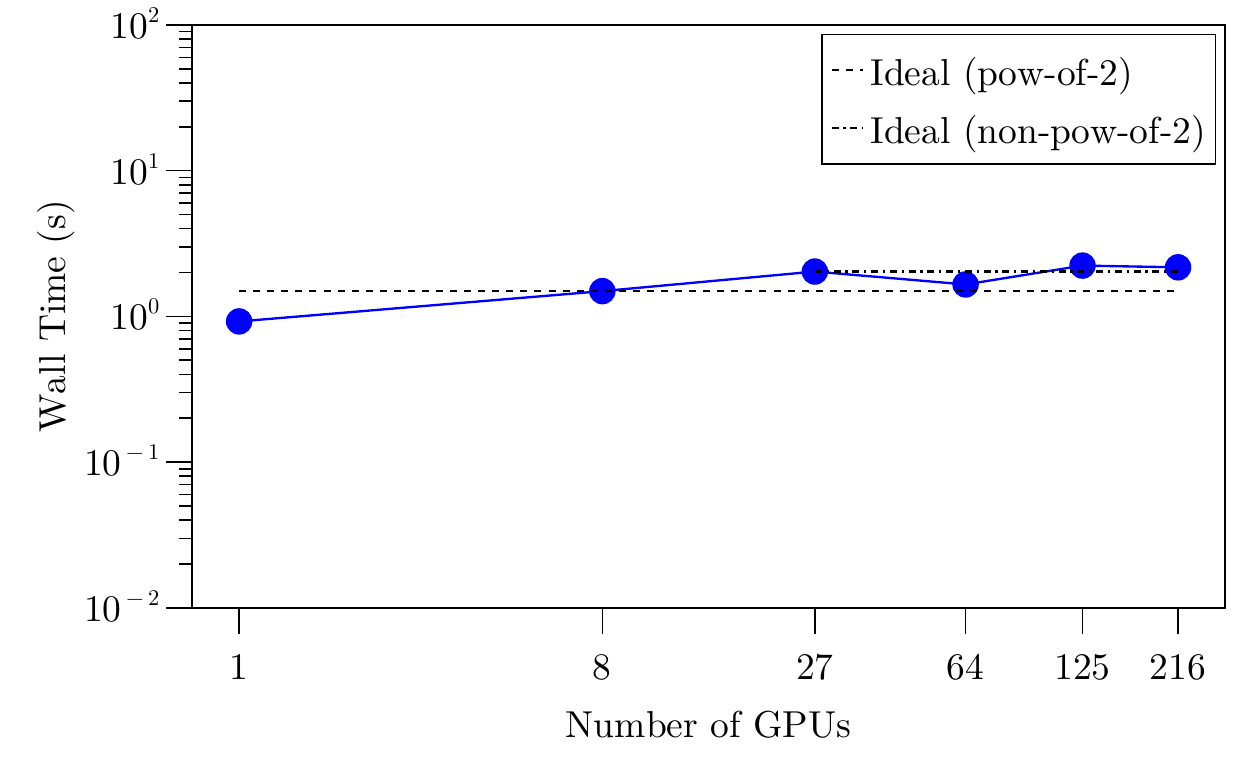}\\
(a)\\[0.5em]
\includegraphics[width=0.80\textwidth]{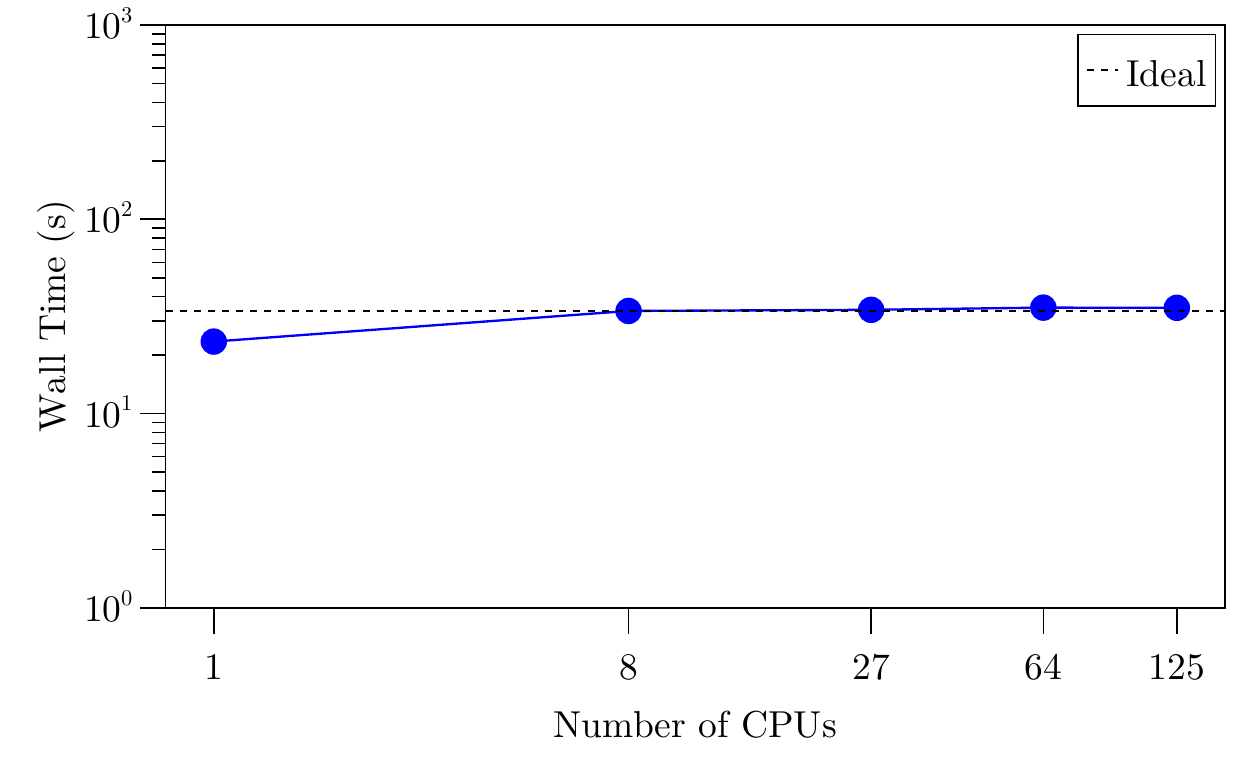}\\
(b)
\caption{Wall-clock time per time-integration step as a function of the number of devices for a fixed
  problem size per device. (a) GPU execution; (b) CPU execution.
  The black horizontal lines indicate ideal weak scaling.}
\label{fig:weak-scaling}
\end{figure}

Figure~\ref{fig:weak-scaling} shows weak scaling results for GPU and CPU.
The problem size per device was set to $N^3 = 750^3$ grids.
Since TSUBAME4.0 has 4 GPUs and 2 CPUs per node, scaling was measured up to 216 GPUs
and 108 CPUs due to node count constraints.
Weak scaling efficiency is defined as $E_w = T_1 / T_N$, where $T_1$ is the execution
time on 1 device and $T_N$ is the execution time on $N$ devices.
For the scaling measurements on CPUs, 8 CPUs were used as the baseline.
For the scaling measurements on GPUs, the baseline device count was 8 for power-of-two
counts (8 and 64) and 27 for non-power-of-two counts (27, 125, and 216).

\begin{figure}[t]
  \begin{minipage}[b]{0.65\textwidth}
    \centering
    \includegraphics[width=\textwidth]{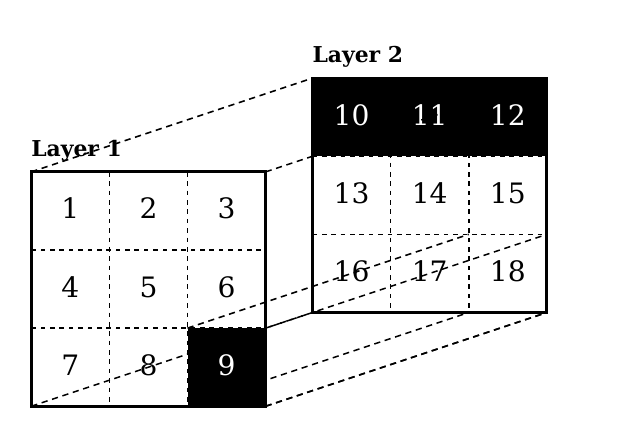}
    \subcaption{}
    \label{fig:comm-patterns-a}
  \end{minipage}
  \hfill
  \begin{minipage}[b]{0.32\textwidth}
    \centering
    \includegraphics[width=\textwidth]{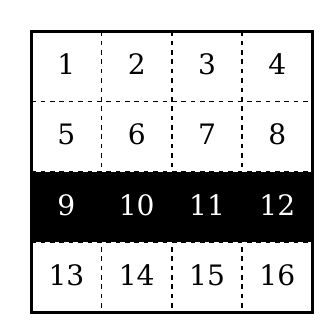}
    \subcaption{}
    \label{fig:comm-patterns-b}
  \end{minipage}
  \caption{Communication patterns for (a) a non-power-of-two $3\times3\times3$ GPU
    topology and (b) a power-of-two $4\times4\times4$ GPU topology.
    GPUs within the same node are shown in black.
    In (a), certain GPUs such as GPU~9 have no intra-node neighbors in any direction.
    In (b), at least one direction of communication remains intra-node for all GPUs.}
  \label{fig:comm-patterns}
\end{figure}

The execution time of GPUs showed two distinct trends depending on whether the device
count was a power of two.
Power-of-two counts achieved approximately 25\% shorter execution time than
non-power-of-two counts.
This difference can be attributed to the node assignment of processes.
Each node contains 4 GPUs.
Therefore, GPUs 1--4, 5--8, 9--12, and so on are assigned to the same node.
Figure~\ref{fig:comm-patterns} illustrates the difference in inter-node communication
patterns between a power-of-two and a non-power-of-two GPU configuration.
In the non-power-of-two case (Fig.~\ref{fig:comm-patterns}a), certain
processes---such as GPU~9---have neighbors across node boundaries in all directions,
forcing all three communications to be inter-node.
In the power-of-two case (Fig.~\ref{fig:comm-patterns}b), some directions of
communication remain intra-node.
Since execution time is bottlenecked by the slowest process, configurations in which all
three communications are inter-node result in a significant performance penalty.
Therefore, power-of-two counts used 8 GPUs as the baseline, and non-power-of-two counts
used 27 GPUs.

Figure~\ref{fig:weak-scaling} confirms that weak scaling efficiency was maintained above
90\% for GPUs (up to 216) for both power-of-two and non-power-of-two configurations,
and above 95\% for CPUs (up to 108).

\subsubsection{GPU and CPU Performance Comparison}\label{sec:gpu-vs-cpu-comparison}

Figure~\ref{tab:gpu-cpu-comparison} shows the performance comparison between GPU and CPU
executions using the same number of nodes on TSUBAME4.0.
Since each node has 4 GPUs and 2 CPUs, the GPU and CPU counts differ.
Because the process and thread balance optimization was not applied during the scaling
experiments, the CPU execution time was estimated by multiplying the measured
optimization efficiency of 2.7 (Sect.~\ref{sec:scaling}).

LCS.jl achieved a maximum speedup of $18.0\times$ on GPUs over CPUs.
This demonstrated that the single-source and multi-platform design can achieve an
acceleration ratio comparable to GPU-dedicated implementations.

\begin{table}[t]
\caption{Wall-clock time per time-integration step for GPU and CPU execution on the same number of nodes
  on TSUBAME4.0. CPU execution time is estimated by multiplying the measured
  optimization efficiency of 2.7 (see Sect.~\ref{sec:scaling}).}
\label{tab:gpu-cpu-comparison}
\begin{tabular}{cccccccc}
\tophline
$N^3$ & $N_p$ & Nodes & GPUs & GPU Time [s] & CPUs & CPU Time [s] (Optimized) & Speedup \\
\middlehline
$1500^3$ & $750^3$  & 2  & 8   & 1.491 & 4   & 20.4  & $13.7\times$ \\
$1500^3$ & $750^3$  & 4  & 16  & 0.779 & 8   & 12.5  & $16.0\times$ \\
$1500^3$ & $750^3$  & 8  & 32  & 0.439 & 16  & 7.19  & $16.4\times$ \\
$1500^3$ & $750^3$  & 16 & 64  & 0.249 & 32  & 4.47  & $18.0\times$ \\
$3000^3$ & $1500^3$ & 16 & 64  & 1.657 & 32  & 23.3  & $14.1\times$ \\
$3000^3$ & $1500^3$ & 32 & 128 & 0.826 & 64  & 13.0  & $15.7\times$ \\
$3000^3$ & $1500^3$ & 64 & 256 & 0.484 & 128 & 7.50  & $15.5\times$ \\
\bottomhline
\end{tabular}
\belowtable{}
\end{table}

\subsubsection{Heterogeneous Execution}\label{sec:heterogeneous-execution}

\begin{table}[t]
\caption{Wall-clock time per time-integration step for CPU-only and Heterogeneous configurations
  ($N^3 = 256^3$, $N_p = 256^3$). ``Reduction from CPU-only'' denotes the percentage
  reduction in wall-clock time relative to the CPU-only configuration.}
\label{tab:heterogeneous}
\begin{tabular}{lcccc}
\tophline
\textbf{Configuration} & \textbf{CPU Cores} & \textbf{GPUs} & \textbf{Time [s]} & \textbf{Reduction from CPU-only} \\
\middlehline
CPU-only      & 32 & ---  & 80.0 & ---   \\
Heterogeneous & 32 & 1    & 22.3 & 0.72 \\
\bottomhline
\end{tabular}
\belowtable{}
\end{table}

LCS.jl supports heterogeneous execution, in which fluid and particle time-stepping and
particle statistics computation are assigned to different computational resources, all
from a single codebase.
A configuration was evaluated in which time-stepping runs on CPUs and particle
statistics computation runs on a GPU.

Computation was performed on a workstation, assuming an environment in which a GPU is
installed as an auxiliary device.
The CPU was 2~$\times$ Intel Xeon Gold 6326 and the GPU was an NVIDIA RTX A6000
(48\,GB VRAM); this configuration represents a typical computing environment in which
assigning one high-performance GPU per CPU is not feasible. 
Consequently, the working dataset often exceeds GPU memory capacity, resulting in degraded performance.

Table~\ref{tab:heterogeneous} shows the execution time per time-integration step for $256^3$ grids and
$256^3$ particles.
Compared to the CPU-only configuration, the Heterogeneous configuration reduced
execution time by approximately 72\%.
The configuration change was realized by switching only the resource assignment, without
modifying the algorithm or physical model.

This trial heterogeneous execution, delegating only statistics computation to a GPU, was
effective even in configurations where the GPU was not the primary compute device.
LCS.jl enables flexible utilization of computational resources in scenarios where budget
or power constraints prevent assigning high-performance GPUs to all processes, or where
GPUs are added incrementally to an existing CPU cluster.


\conclusions

We developed LCS.jl (Lagrangian Cloud Simulator in Julia), a single-source and
multi-platform multiphase turbulence simulation model implemented in Julia language and
KernelAbstractions.jl.

Validation results confirmed that the present fluid and particle statistics agree well
with those obtained in prior studies.
This confirms the validity of both the fluid solver and the particle tracking
implementation.

The particle communication cost in the GPU-native implementation based on prefix-scan
was 8.0 times faster than the theoretical lower bound of the CPU-delegated approach.
The GPU-native implementation reduced the particle communication cost from approximately
78\% (CPU-delegated) to 10\% of total execution time.
This reduction eliminated the communication bottleneck that would otherwise negate the
computational advantage of GPU execution, enabling effective utilization of GPU
performance in large-scale multiphase flow simulations.

We confirmed that the present LCS.jl achieves computational performance equivalent to
the Fortran implementation in computations using many processes.
This demonstrated that the single-source and multi-platform design does not sacrifice
computational performance.
The performance results confirmed that large-scale DNS computations can be performed
efficiently on any platform.
For example, for GPUs, strong scaling efficiency was maintained above 85\% (up to 256
GPUs) and weak scaling efficiency above 90\% (up to 216 GPUs).
For CPUs, strong scaling efficiency was maintained above 70\% (up to 128 CPUs), and
weak scaling efficiency above 95\% (up to 108 CPUs).

LCS.jl achieved a maximum speedup of $18.0\times$ on GPUs over CPUs.
This demonstrated that the single-source and multi-platform design achieves an
acceleration ratio comparable to GPU-dedicated implementations.
A trial heterogeneous execution, delegating only statistics computation to a GPU,
achieved a 72\% reduction in execution time compared to the CPU-only configuration even
in configurations where the GPU was not the primary compute device.
This means that LCS.jl enables flexible utilization of computational resources in
scenarios where budget or power constraints prevent assigning high-performance GPUs to
all processes, or where GPUs are added incrementally to an existing CPU cluster.

The present LCS.jl is not merely a GPU-ported code, but a multiphase turbulence
simulation platform that achieves both portability and scalability across a variety of
computational resource configurations.
This design philosophy can provide a guideline for building high-performance
computational models that can adapt to future architectural changes and heterogeneous
computing environments.

\noappendix

\codeavailability{
The source code of LCS.jl is publicly available on GitHub at
  \url{https://github.com/0samuraiE/LCS.jl} and archived on Zenodo at
  \url{https://doi.org/10.5281/zenodo.19515891}.
}

\authorcontribution{T.\ Tominaga developed the model code, performed the simulations,
and prepared the manuscript.
R.\ Onishi supervised the research and contributed to manuscript revision.}

\competinginterests{The authors declare that they have no conflict of interest.}

\begin{acknowledgements}
This work used computational resources TSUBAME4.0 supercomputer provided by Tokyo
Institute of Technology through Joint Usage/Research Center for Interdisciplinary
Large-scale Information Infrastructures and High Performance Computing Infrastructure in
Japan (Project ID: jh240041).
\end{acknowledgements}

\bibliographystyle{copernicus}
\bibliography{refs}

\end{document}